\documentclass[epj,referee]{svjour}
\usepackage{graphicx}
\usepackage{dcolumn}
\usepackage{amsmath}
\usepackage{amssymb}
\usepackage[latin1]{inputenc}
\usepackage{graphics}

\newcommand{\rmd}{{\rm d}}
\newcommand{\rmi}{{\rm i}}

\newcommand{\Dt}{{\Delta t}}

\newcommand{\ts}{{t_{\rm s}}}
\newcommand{\tOc}{{t_{0\rm c}}}
\newcommand{\tOm}{{t_{0\rm m}}}

\newcommand{\chideuxm}{{\chi^2_{\rm m}}}
\newcommand{\Dks}{{D_{\rm KS}}}
\newcommand{\Dtks}{{\Delta t_{\rm KS}}}
\newcommand{\SOm}{S_{0\rm m}}
\newcommand{\sOm}{s_{0\rm m}}
\newcommand{\rOm}{r_{0\rm m}}
\newcommand{\Sc}{{S_{\rm c}}}
\begin{document}

\title{Kolmogorov-Smirnov method for the determination of signal time-shifts.}

\author{ P. Désesquelles\inst{1} \and T.M.H. Ha\inst{1} \and A. Korichi\inst{1}\and  F. Le Blanc\inst{2} \and C.M. Petrache\inst{2}}                    
% Do not remove
%
\mail{Pierre.Desesquelles@in2p3.fr}%
\institute{CSNSM CNRS/IN2P3 and Université Paris Sud 11, 15 rue G. Clémenceau, 91405 Orsay, France \and IPNO CNRS/IN2P3 and Université Paris Sud 11, 15 rue G. Clémenceau, 91405 Orsay, France\\
\newline
On behalf of the AGATA Collaboration}
\date{Received: date / Revised version: date}
% The correct dates will be entered by Springer
%
\abstract{
A new method for the determination of electric signal time-shifts is introduced. As the Kolmogorov-Smirnov test, it is based on the comparison of the cumulative distribution functions of the reference signal with the test signal.
This method is very fast and thus well suited for on-line applications. It is robust to noise and its performances in terms of precision are excellent for shift times ranging from a fraction to several sample durations.
\PACS{
      {29.40.Gx}{Tracking and position-sensitive detectors}   \and
      {29.30.Kv}{X- and gamma-ray spectroscopy}   \and
      {07.50.Qx}{Signal processing electronics}
     } % end of PACS codes
} %end of abstract
\maketitle

\section{Introduction}

Modern detection systems depend more and more on embedded intelligence that allows the on-line determination of the characteristics of the impinging particles \cite{Ham1,Lee1,Cha1,Mar2}. In many cases, the full shape of the delivered signals is first digitized and then compared to a basis of reference signals. The basis signals reflect all the possible values of the characteristics of the particles. The properties of the particle actually detected are supposed to be the same as those of the best-matched signal of the basis. However, random time jitters often alter detector signals, which can severely affect their comparison to basis signals even when the time-shift is smaller than the sample duration \cite{Des5}. More generally, signal absolute-time alignment with respect to a reference signal or relative-time alignment (e.g. with respect to the signal set average) are important issues in detector signal processing. In most cases, on-line time-correction techniques, such as constant fraction discrimination, rely on the crossing of a threshold. Due to the signal noise, and in many applications to the digitization, these techniques do not allow time correction shorter than the sample duration (intra-sample correction). The poor-man solution to time jitters consists in a chi-square comparison of the detected signal with different versions of time-shifted reference signals. This solution is often inapplicable as it may severely slow down the acquisition rate. In the following, we propose an algebraic determination of the time-shift based on the Kolmogorov-Smirnov (KS) distance. This method consists in measuring the time distance between the detected signal and a reference signal. It is well adapted to on-line applications since it is very fast and it entails only a small memory space. It will be shown that the resolution of the KS method is excellent, with respect to the usual time alignment methods, for noisy signals and when the time-shift is a few sample durations or shorter. Therefore, for longer time shifts, it can be used to refine a first evaluation obtained by another method.

The method is simple to implement and should be well suited to a large number of applications. However, it is not the case for the measurement of gas chambers drift-times, which could have been an important application. Indeed, the KS method requires well defined reference signals but also that the detector signals be affected by uncorrelated white noise only. Gas chamber signals are altered by strong auto-correlated fluctuations due to the diffusion phenomena in the gas.

In Section 2, we summarize the KS method and show why it is more sensitive to  curve shifts than the usual chi-square method. An illustration of the method, using High Purity Germanium (HPGe) detector signals, is given in Section 3. The conclusions are enumerated in Section 4.

\section{The Kolmogorov-Smirnov test}

\subsection{Description of the method}

\begin{figure}[htbp]
\begin{center}
\includegraphics[width=7.5cm]{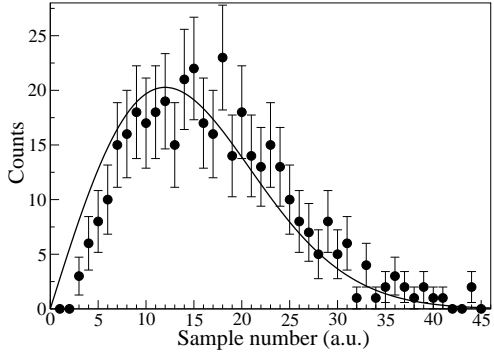}
\caption{Comparison of a generic theoretical distribution (full line) with experimental count numbers (black dots) as a function of the sample number (this figure shows an example made by hand only to illustrate the method, it does not present real data).}
\label{Fig f_theo_exp}
\end{center}
\end{figure}

\begin{figure}[htbp]
\begin{center}
\includegraphics[width=7.5cm]{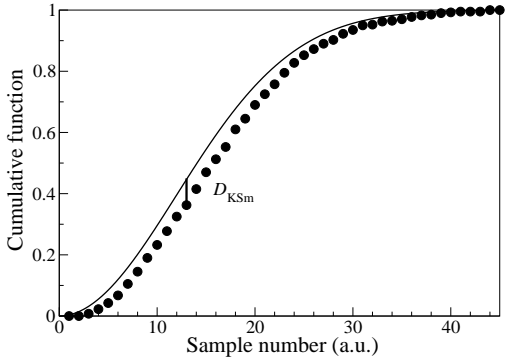}
\caption{Comparison of the cumulative functions of the distributions from Fig.~\ref{Fig f_theo_exp}. The KS distance is the absolute value of the maximum difference between the two.}
\label{Fig frep_theo_exp}
\end{center}
\end{figure}

The KS distance is, as the chi-square, a criterion to qualify the agreement between two distributions. Its implementation is illustrated in Figs. \ref{Fig f_theo_exp} and \ref{Fig frep_theo_exp}. In the first figure a "theoretical" distribution is compared to "experimental" points affected by statistical fluctuations. The KS distance between the two distributions is defined as the maximum distance $\Dks$ between their cumulative distribution functions (see Fig.~\ref{Fig frep_theo_exp}). The cumulative distribution function of a probability density function $f$ is defined as $F(x)=\int_{-\infty}^x f(x')\, \rmd x'$. We use the same term for discrete functions: $F(j)=\sum_{i=1}^j f_{i}$, with $f_{i}=n_{i}/N_{\rm tot}$, where $n_{i}$ is the number of counts in the $i^{\rm th}$ sample and $N_{\rm tot}$ the total number of counts. Cumulative distribution functions are monotonically increasing from zero to one.

\begin{figure}[htbp]
\begin{center}
\includegraphics[width=7.5cm]{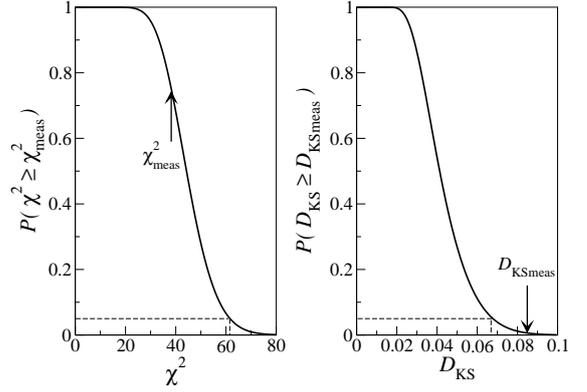}
\caption{Probability that the measured chi-square (left) and measured KS distance  (right) are greater than a given value. The dotted lines correspond to a probability of 5\%. The shape of the chi-square distribution depend only on the number of samples, while the shape of the KS distance distribution depends only on the total number of counts.}
\label{Fig chi2_Dks}
\end{center}
\end{figure}

The so-called chi-square distribution gives the probability that the measured un-normalized chi-square ($\chideuxm$) is superior to a given value. Such a distribution also exists for the KS distance (Fig.~\ref{Fig chi2_Dks}). In our case, if the differences between the theoretical curve of Fig.~\ref{Fig f_theo_exp} and the experimental data are only due to the fluctuations, then the chi-square has a 5\% probability of being larger than 62 and the KS distance has a 5\% probability of being larger than 0.066. If the measured values are greater than about 70 and 0.08 respectively then the theoretical curve is most probably invalidated by the experimental results.

The KS distance and the chi-square are not sensitive to the same types of discrepancies between the distributions under comparison. It is visible from Fig.~\ref{Fig f_theo_exp} that the theoretical curve is shifted to the left with respect to the experimental points. However, the curve crosses most of the error bars. Hence, the measured chi-square has a low value ($\chideuxm = 40.4$) from which one could conclude that both distributions are in agreement. Indeed, as shown in Fig.~\ref{Fig chi2_Dks}, the probability to obtain a chi-square equal or greater than 40.4 is 67\%. However, the KS distance for the same distributions is 0.084, which corresponds to a very low probability of 0.6 \%. This shows that the theoretical curve is most probably invalidated by the data. The KS distance is a test particularly sensitive to the shifts between distributions. In the following, this well established result of probability theory will be applied to determine the time-shifts of detector signals affected by noise and a random time jitter.

\subsection{Application to the determination of random time-shifts}

The cumulative function $S$ is now obtained by integrating the signal $s$ and normalizing the maximum to one:

\begin{equation}
\label{Eq S}
S(t) = \frac{\sum_{t'=1}^t s(t')}{\sum_{t'=1}^{t_{\rm max}} s(t')}
\end{equation}

\noindent Our goal is not to measure precisely the KS distance between the shifted test signal $s$ and the reference signal $s_{0}$, but to measure the time-shift in a fast way. The vertical distance between slightly shifted distributions is large where the slope is large, as can be seen for example in Fig.~\ref{Fig frep_theo_exp}, since, to the first order,  $\Delta S = (\rmd S/\rmd t)\,\Delta t$. The maximum slope of the cumulative distribution corresponds to the maximum of the signal (in the example of Fig.~\ref{Fig f_theo_exp}, $\tOm = 13$). Therefore, the distance between the reference signal cumulative distribution $S_{0}$ and the test signal cumulative distribution  $S$ will be evaluated at $t=\tOm$. 

\begin{figure}[htbp]
\begin{center}
\includegraphics[width=8cm]{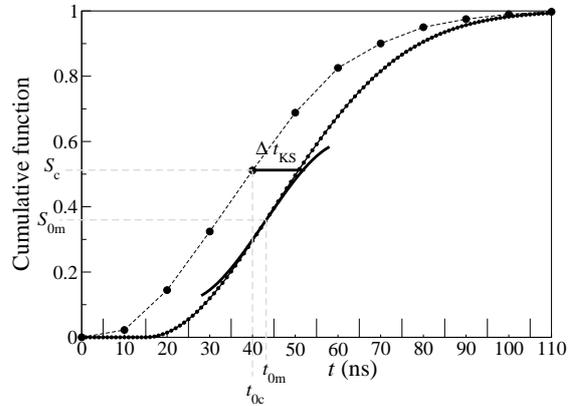}
\caption{Scheme of the determination of the time-shift using the Kolmogorov-Smirnov method. The full line curve is $S_{0}$, the reference signal cumulative function (1~ns time bins) and the dotted curve is $S$, the test signal cumulative function (10~ns time bins). The bold curve is the third order Taylor expansion of the reference signal around $\tOm$. The estimation of the time-shift is the distance between the dotted and the bold curves at $S=\Sc$.}
\label{Fig schema}
\end{center}
\end{figure}

Noting $\tOc$ the central time of the sample which contains the reference signal maximum (see Fig.~\ref{Fig schema}), and $\ts$ the sample duration:

\begin{equation}
\label{Eq tm}
\tOc - \frac{\ts}{2}\ \leq \tOm < \tOc + \frac{\ts}{2}\,,
\end{equation}

\noindent we have:

\begin{equation}
\label{Eq S(tm)}
S(\tOc) = S_{0}(\tOc+\Dt)\,,
\end{equation}

\noindent where $\Dt$ is the time-shift between the test and the reference cumulative distributions. 

The value of the reference cumulative distribution function can be evaluated using its Taylor polynomial \cite{Des4}:

\begin{eqnarray}
\label{Eq S0(tm-Dt)}
S_{0}(\tOc+\Dt) &=& S_{0}(\tOm)  \\
\nonumber
&+&(\tOc+\Dt-\tOm)\ \left.\frac{\rmd S_{0}}{\rmd t}\right|_{\tOm} \\
\nonumber
&+& \frac{1}{2}\,(\tOc+\Dt-\tOm)^2\ \left.\frac{\rmd^2 S_{0}}{\rmd t^2}\right|_{\tOm} \\
\nonumber
&+& \frac{1}{6}\,(\tOc+\Dt-\tOm)^3\ \left.\frac{\rmd^3 S_{0}}{\rmd t^3}\right|_{\tOm}\\
\nonumber
&+& \ldots 
\end{eqnarray}

\noindent The cumulative distribution function is obtained by integrating the signal (\ref{Eq S}) thus, its first derivative is simply the signal itself: $(\rmd S_{0}/\rmd t)(\tOm) = s_{0}(\tOm)$. The second derivative is equal to zero as it corresponds to the slope of the signal when it reaches its maximum. The third derivative can be estimated by:

\begin{equation}
\label{Eq r0}
\rOm  = \frac{s_{0}(\tOm+2)-2\,s_{0}(\tOm)+s_{0}(\tOm-2)}{4\,\ts^2}\,,
\end{equation}

\noindent where $\ts$ is the sample duration (1~ns in our case). Using Eqs. (\ref{Eq S(tm)}), (\ref{Eq S0(tm-Dt)}) and (\ref{Eq r0}), and simplifying the notations, one obtains:

\begin{eqnarray}
\label{Eq Sm}
\Sc &=& \SOm  \\
\nonumber
&+&(\tOc+\Dtks-\tOm)\   \sOm\\
\nonumber
&+& \frac{1}{6}\ (\tOc+\Dtks-\tOm)^3\  \rOm\,.
\end{eqnarray}

\noindent In this equation, all the constants $\tOm$, $\tOc$, $\SOm$, $\sOm$, $\rOm$ depend only on the reference signal and thus, can be pre-calculated off-line. Only these constants have to be saved in the computer memory, instead of the full signals. The value of $\Sc$ is calculated by integrating the test signal. Hence, Eq. (\ref{Eq Sm}) is a completely defined third order equation in $\Dtks$. The first order solution (which is also the second order solution since $(\rmd^2 S_{0}/\rmd t^2)(\tOm)=0$) reads:

\begin{equation}
\label{Eq Dto2}
\Dtks = \frac{\Sc-\SOm}{\sOm}+\tOm-\tOc\,.
\end{equation}

\noindent The derivative of the resulting time-shift with respect to the test signal is $(\partial \Dtks/\partial \Sc) = 1/\sOm$. As $\sOm$ is the maximum of the signal, the error propagation from the statistical fluctuation on the cumulative distribution to the time-shift evaluation is minimized.

The complete third order solution reads:

\begin{equation}
\label{Eq Dto3}
\Dtks =\frac{(1-\rmi\sqrt{3})\,\sOm}{\alpha}-\frac{(1+\rmi\sqrt{3})\,\alpha}{2\,\rOm}
+\tOm-\tOc\,,
\end{equation}

\noindent with:

\begin{eqnarray}
\label{Eq alpha}
\nonumber
\alpha &=& \big[\sqrt{\rOm^3(8\,\sOm^3+9\,\rOm(\Sc-\SOm)^2)}\\
&+&3\,\rOm^2(\Sc-\SOm)\big]^{\frac{1}{3}}\,.
\end{eqnarray}

\section{Results}
\subsection{The set of test signals}

\begin{figure}[htbp]
\begin{center}
\includegraphics[width=7.8cm]{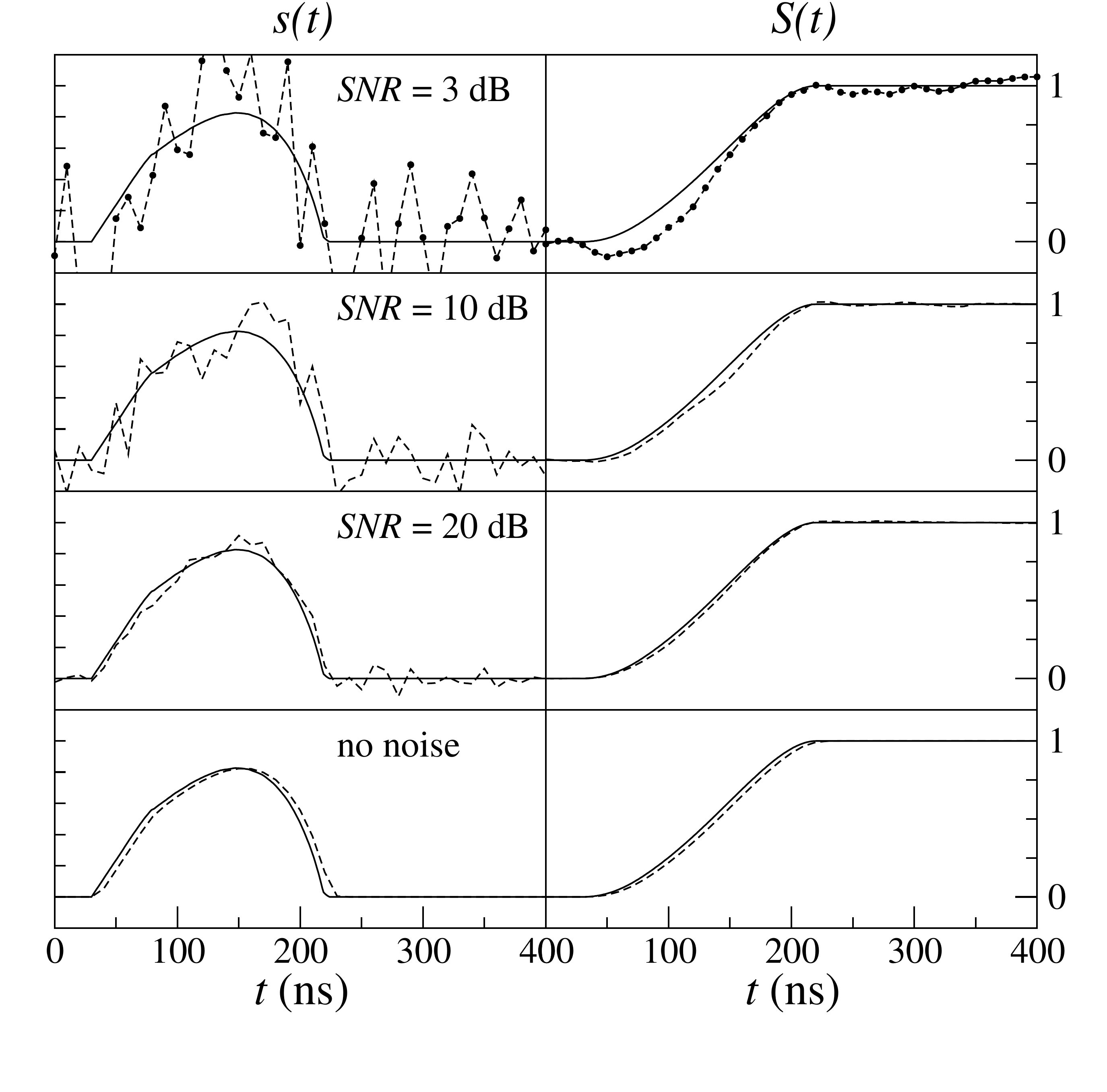}
\caption{One example of signal (left column) and the corresponding cumulative distribution function (right column). The reference signal (1 ns samples) is represented in full line and the test signals (10 ns samples) are represented in dotted line. The test signals are shifted by 10 ns with respect to the reference signal. The amount of noise is indicated for each line.}
\label{Fig graph42}
\end{center}
\end{figure}

In order to illustrate the method, we apply it to large sets of simulated signals with different amounts of noise and different time-shifts. These signals were generated using the MGS code \cite{Med1,Nel1,Ola1}, that simulates HPGe signals \cite{Des3}. A large variety of  rise times and pulse shapes is obtained. The sample duration of the simulated signal is 1~ns. As the experimental signals, they are grouped into 10~ns samples. Eventually, white noise is added to the signals. The reference signals are the 1~ns plane signals and the test signals are the 10~ns noisy and shifted signals (respectively in full and dotted lines in the example of Fig.~\ref{Fig graph42}). The KS method applies to positive distributions, thus, in the case of non-positive electric pulses, their absolute values must be considered. In the specific case of segmented HPGe detectors, there are two types of signals: transient signals, and signals from the hit segments which are charge-integrated, so that their shape is similar to a cumulative distribution function. In the latter case, the KS method can be applied directly to the signal, after normalization of its amplitude to 1. Unfortunately, in this case, one does not benefit from the smoothing property of signal integration.

In the following, we do not consider the problem of finding the basis reference signal that corresponds to the test signal. This problem is crucial in many applications. For discussion of the methods to determine the right reference signal, in the case of HPGe detectors, see \cite{Des5,Des4,Ola1,Des3,Dox1} and references therein.

\subsection{KS method}

\begin{figure}[htbp]
\begin{center}
\includegraphics[width=8cm]{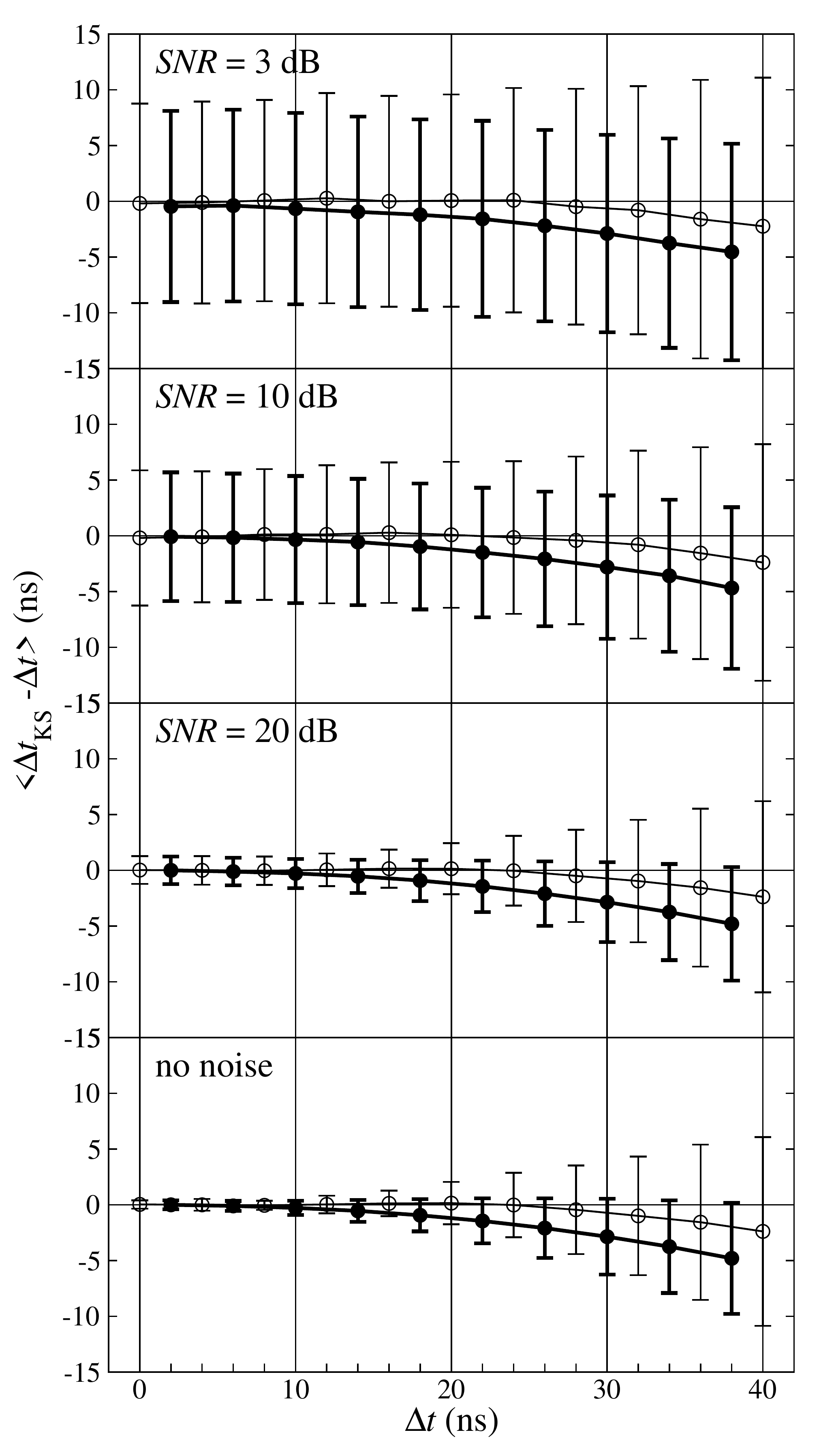}
\caption{Error on the time-shift determination using the Kolmogorov-Smirnov method. The black dots correspond to the first order correction and the open dots correspond to the third order correction. The error bars give the one-sigma width (one thousand test signals per dot).}
\label{Fig graph41}
\end{center}
\end{figure}

The application of the first order and third order corrections to noisy and shifted test signals is illustrated in Fig.~\ref{Fig graph41}. The plots show the error on the correction as a function of the imposed time-shift for different values of the signal-to-noise ratio. The down-bendings of the curves are due to the truncation of the third and fourth order terms in the Taylor expansion.

Inside the first 10~ns sample, both corrections are unbiased whatever the noise amount. The third order correction is still unbiased at the end of the second bin. Beyond the second bin, a small negative drift is visible (the time-shift is slightly underestimated). The error bars show that the method reaches its limit ($\Dtks-\Dt\approx 10$~ns) at $SNR = 3$~dB which is a usual minimum value of the signal-to-noise ratio when extracting information from a noisy signal.

\subsection{Comparison with other methods}

\begin{figure}[htbp]
\begin{center}
\includegraphics[width=8cm]{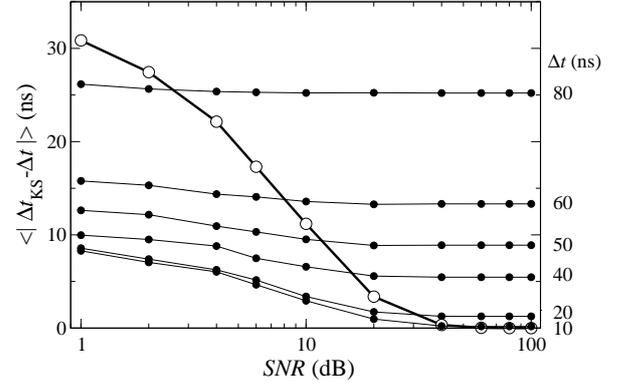}
\caption{Comparison of the average absolute value of the error on the time-shift determination for the first order KS method (black dots) and the method of the centroid (open dots) as a function of the signal-to-noise ratio. The simulated time-shift ranges from 10 ns to 80 ns. The results obtained with the centroid method are almost independent of the imposed time-shift. The sample duration is 10~ns}
\label{Fig moy_abs_et_cen_o1}
\end{center}
\end{figure}

A large number of methods for the determination of time-shifts are available. Electronic or software CFD \cite{Jac1} are fast but very sensitive to noise and, except for very high signal-to-noise ratios, not adapted to intra-sample correction. The chi-square or residue methods consist in the extensive comparison of the test signal with different, regularly time-shifted, versions of the reference signal \cite{Dox1}. This procedures allow a comparable time resolution but are about two orders of magnitude slower than the solution we propose and entails the handling of bases of shifted reference signals that may be very large. Another popular technic consists in calculating the centroid time \cite{Rus1} of the detected pulse : $t_{\rm cen} = \sum_t t\,s(t)/\sum_t s(t)$. The computation time is the same as for the KS method (Eqs. (\ref{Eq S}) and (\ref{Eq Dto2})). The comparison of the absolute value of the error fot both methods is presented in Fig.~\ref{Fig moy_abs_et_cen_o1}. The KS method appears to be less sensitive to the noise and its resolution is much better for time-shifts smaller than a few sample durations.

\section{Conclusion}

We have presented a method, based on the Kolmogorov-Smirnov test, to correct signals corrupted by time jitters. This method appears to be very efficient for the precise correction of time-shifts up to a few bin times. For $SNR$ larger than 20 dB, the bias for intra-sample correction is less than one tenth of the sample duration. The good precision of the corrections is connected to two features of the method. On the one hand, the time-shift is calculated from the cumulative distribution function of the signal (so that the noise fluctuations are lowered by the integration, see Fig.~\ref{Fig graph42}), and, on the other hand, it is evaluated where the pulse amplitude is maximum so that the error propagation is minimized. 

The first order correction is extremely fast since it entails only the integration of the signal and the calculation of a first degree polynomial, Eq. (\ref{Eq Dto2}). The third order correction is non-linear complex, Eq. (\ref{Eq Dto3}). Its bias is smaller than in the first order case, but, as it includes the calculation of higher powers, the uncertainties are larger. 

In our AGATA example, the reference signals are simulated. The method can also be used for signal alignment. In this case, the reference signal could simply be the mean of the measured signal set.

Finally, due to its simplicity, the method is versatile whereas it is probably not suited to any type of signals (for example when the signals cannot be corrected from baseline / DC offsets). In the case of HPGe detector signals, its reliability is comparable with other methods for long time-shifts, while, in the range of one tenth to a few sample durations the results, in terms of resolution, robustness to the noise and computing time, are excellent.

\end{document}